\documentclass[prb,preprint,showpacs,preprintnumbers,amsmath,amssymb,floatfix]{revtex4}
\usepackage{graphicx}% Include figure files
\usepackage{epsfig}
\usepackage{dcolumn}% Align table columns on decimal point
\usepackage{bm}% bold math
\usepackage{placeins}
\usepackage{flafter}
\usepackage{color}

\begin{document}

\newcommand{\pnma}[1]{Pnma-{\textbf #1}}

\title{ IR-active optical phonons in \pnma{1}, \pnma{2} and R$\bar 3$c phases of LaMnO$_{3+\delta}$}

\author{I.~S.~Smirnova}
\email{smirnova@issp.ac.ru}
\author{A.~V.~Bazhenov}
\author{T.~N.~Fursova}
\author{A.~V.~Dubovitskii}
\author{L.~S.~Uspenskaya}
\author{M.~Yu.~Maksimuk}
\affiliation{ Institute of Solid State Physics, Russian Academy of Sciences, 142432, Chernogolovka, Moscow distr.}

\begin{abstract}

Infrared-active phonons in LaMnO$_{3+\delta}$ were studied by means of the reflection and transmission spectroscopy from 50 to 800 cm$^{-1}$ at room temperature. Powder and ceramic samples of the phases of \pnma{2} ($\delta=0.02$), \pnma{1} ($\delta=0.08$), and R$\bar 3$c ($\delta=0.15$) were investigated. Besides, energies of the dipole-active phonons in \pnma{2}, \pnma{1} phases were obtained by lattice-dynamics calculations. The transformations of IR-active phonons with the increase of $\delta$ in the sequence of \pnma{2}, \pnma{1}, R$\bar 3$c are discussed.

\end{abstract}

\pacs{61.50.Ah, 78.30.-j}

\maketitle

\section{Introduction}

Electrical and magnetic properties of perovskitelike compounds R$_x$A$_y$MO$_{3+\delta}$ (R = rare earth; A = Ca, Sr, Ba, Pb; M = Cu, Mn, Ti, V) can drastically change with varying $x, y$. In last decades such materials have been intensively investigated. In 1986 superconductivity with $T_c$=30~K was found in La$_{2-x}$Ba$_x$CuO$_{4+\delta}$.[\onlinecite{Muller}] Using of some other transition metals M can result in the compounds (La$_{1-x}$A$_x$MnO$_{3+\delta}$, for example) with ``colossal" magnetoresistance.\cite{vonHem,Chahara,Jin} Cuprates and manganites possess many common features: the crystal structure (close to the perovskite) and strong electron--electron, electron--phonon, and exchange interaction. With $x=0, \delta=0$ they are antiferromagnetic dielectrics at low temperatures. An increase of $x$ results in a dielectric-metal transition.\cite{Loktev}

Some excess oxygen in La$_2$CuO$_{4+\delta}$ brings about the same transformation of the electronic spectrum that results from the partial substitution of La by an alkali earth, the transformation going up to the superconducting phase.\cite{Zakharov} Similarities between cuprates and manganites stimulated studies of the influence of excess oxygen on the electron and phonon spectra of LaMnO$_{3+\delta}$. 

It's well known that the crystal structure of both LaMnO$_{3+\delta}$ and La$_{1-x}$A$_x$MnO$_3$ is orthorhombic at $\delta<0.1$, $x<0.2$ and $T<500$~K. An increase of $\delta$ and $x$ results in a rhombohedral phase R$\bar 3$c. \cite{Topfer,Kavano,Urushibara} In any case the crystal is insulating and paramagnetic above 200--300~K. With decreasing temperature the R$\bar 3$c phase transforms into an orthorhombic phase, insulating and ferromagnetic at $0.11<\delta<0.14$, metallic and ferromagnetic at $\delta>0.14$.\cite{Topfer} Two orthorhombic phases of LaMnO$_{3+\delta}$ have been found.\cite{Mitchell,Hauback,Huang,Topfer} They were denoted as either \pnma{1}, \pnma{2} (Ref. \onlinecite{Huang}) or $O$, $O^\prime$ (Ref. \onlinecite{Topfer}). The first one, \pnma{2} ($O^\prime$), is an insulating antiferromagnet at low temperatures and exists at small $\delta$; the second one, \pnma{2} ($O$), is an insulating ferromagnet at low temperatures and exists at larger $\delta$.

Orthorhombic phases can belong to different space groups (the orthorhombic phase of the La$_2$CuO$_4$, for example, belongs to the Cmca space group). To emphasize that both $O^\prime$ and $O$ phases of LaMnO$_{3+\delta}$ belong to the same space group Pnma we, following Ref. \onlinecite{Huang}, use the notation \pnma{2}, \pnma{1}. Unfortunately, this notation does not show the local symmetries of the atoms or the Wyckoff positions, which are subgroups of the point group $D_{2h}$. It's the local symmetry that determines the number of modes in every irreducible representation.

The purpose of the present study was to examine the spectra of dipole-active optical phonons in \pnma{2}, \pnma{1} and R$\bar 3$c phases. Especially, we paid attention to transformations that phonon states undergo upon transitions from the phase \pnma{2} to \pnma{1} and then to R$\bar 3$c, which are induced by a high-temperature treatment. Optical phonons in the \pnma{2} and R$\bar 3$c phases were measured in Refs. \onlinecite{Iliev1,Iliev2} (\pnma{2}, Raman); \onlinecite{Paolone, Jung} (\pnma{2}, IR); \onlinecite{Abrashev,Iliev2,Iliev3} (R$\bar 3$c, Raman); and \onlinecite{Abrashev,Kebin} (R$\bar 3$c, IR). In the present study, we focus on the IR spectrum of the \pnma{1} phase of LaMnO$_{3+\delta}$. To our knowledge, there are no data on either IR or Raman spectra of this phase at the moment.

The \pnma{2}, \pnma{1} phases are isostructural, so the number of phonon modes should be the same in both cases. However, the number of IR-active modes observed experimentally in the spectra of the \pnma{1} phase is smaller than that for the \pnma{2} phase. In the R$\bar 3$c phase an experiment shows more modes than group theory predicts for the R$\bar 3$c symmetry.

\section{Crystal structure of L$\textrm{a}$M$\textrm{n}$O$_{3+\delta}$ phases}

\begin{figure}
\includegraphics[width=\textwidth,height=\textwidth,keepaspectratio=true]{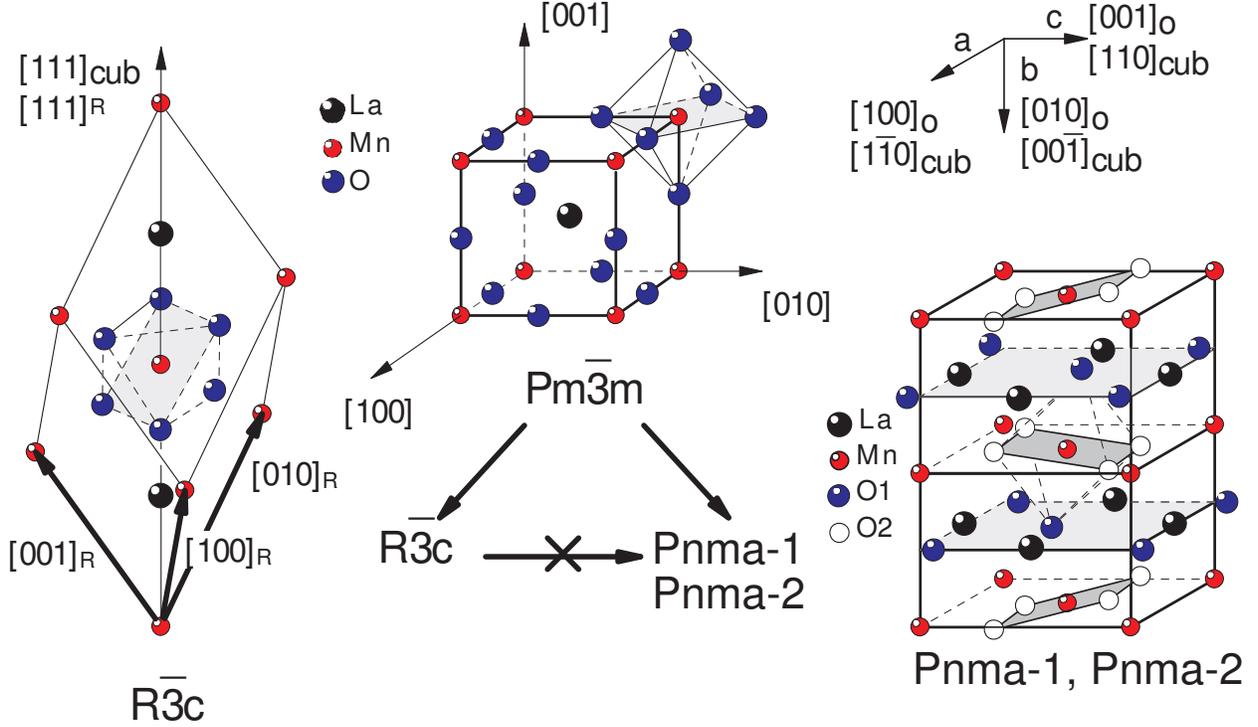}
\caption{\label{fig:fig1} Crystal structure of the R$\bar 3$c (left), Pm$\bar 3$m (centre) and Pnma (right) phases of LaMnO$_{3+\delta}$.}
\end{figure}

Since phonon modes are closely related to the crystal lattice symmetry, let us summarize some well known data about crystal structure of four LaMnO$_{3+\delta}$ phases. The structure of the parent cubic phase Pm$\bar 3$m is shown in the centre of Fig. \ref{fig:fig1}. At ambient pressure, this phase exists at temperatures above 870~K. At room temperature there exist three phases: orthorhombic \pnma{2}, \pnma{1} and trigonal (rhombohedral) R$\bar 3$c.\cite{Elemans,Huang,Mitchell,Hauback}

X-ray analisys shows the following:
\begin{itemize}
\item In the \pnma{2} phase the positions of O2 oxygens (see Fig. \ref{fig:fig1}) deviate considerably from those in the cubic phase. The oxygen octahedra are strongly distorted, particularly in Mn--O2 plane, the Mn--O2 distances differ from each other (1.90 and 2.17 \AA).
\item In the \pnma{1} phase the positions of O2 oxygens slightly deviate from that in the cubic phase, the oxygen octahedra are slightly distorted, the Mn--O2 distances being close to each other.
\item In comparison with the cubic phase, in both orthorhombic phases the oxygen octahedra are rotated around [010] (cubic) axis by nearly the same angle (the difference is 1--3$^\circ$).
\end{itemize}

To distinguish between the \pnma{1} and \pnma{2} phases experimentally, it is sufficient to determine the dimensions of the unit cell: $a, b, c$. In the \pnma{2} phase $a>c$ and $a-c\approx0.2$ \AA, in the \pnma{1} phase $a<c$ and $c-a$ is 0.04--0.08 \AA.

In all considered phases, Mn atoms occupy symmetry-equivalent positions and their time-average charges must be the same. Mn$^{+4}$ should be defects chaotically distributed in the sample volume. Symmetry forbids any long-range charge ordering in these phases. Such ordering may occur only if the symmetry is lowered. 

In contrast to the cubic phase and the trigonal phase, the orthorhombic phases contain two types of inequivalent oxygen atoms. Therefore, these oxygen atoms can have different charges and different amplitudes of displacements in the normal vibration modes. All the six oxygen atoms in the unit cell of the R$\bar 3$c phase are symmetry-equivalent, therefore their scalar parameters, in particular their charges, should be equal. 

Arrows in the centre of Fig. \ref{fig:fig1} show that the point group $D_{3d}$ of the R$\bar 3$c phase and the point group $D_{2h}$ of the Pnma phases are subgroups of the O$_h$ point group of the Pm$\bar 3$m phase and corresponding phase transitions of the second kind are allowed. The crossed arrow in Fig. \ref{fig:fig1} shows that $D_{2h}$  is not a subgroup of $D_{3d}$. As a result, phase transitions of the second kind from the R$\bar 3$c phase to the \pnma{1}, \pnma{2} phases are forbidden. Such phase transitions can be possible only through an increase of symmetry, i. e., through the intermediate cubic phase, which exists at high temperatures.

\section{Experimental}

LaMnO$_{3+\delta}$ was prepared from La$_2$MnO$_3$, La(CO$_3$)$_3$$\cdot$6H$_2$O and Mn$_2$O$_3$. The stoichiometric mixture of source materials was powdered in a ball planetary mill, after that it was calcined at $900^\circ$C, and then it was powdered once again. The main synthesis was conducted at $1100^\circ$C during 10--20 hours. $\delta$ was measured by iodometric titration of the Mn$^{+3}$, Mn$^{+4}$ ions.

It is known that the \pnma{2} phase can be transformed to the \pnma{1} phase by annealing in air. Upon further annealing in oxygen, the \pnma{1} phase transforms into the $R\bar 3$c phase.\cite{Topfer} In Ref. \onlinecite{Iliev1} the \pnma{2} phase was obtained by heating of the R$\bar 3$c phase in N$_2$ atmosphere at $900^\circ$C. We realized the reversible sequence of transformations: R$\bar 3$c $\Leftrightarrow$ \pnma{1} $\Leftrightarrow$ \pnma{2}. First, we kept LaMnO$_{3+\delta}$ powder at $600^\circ$C during 10 hours, then different speeds of cooling resulted in different phases. For the measurements of the IR reflection spectra, ceramic pellets of the \pnma{1}, R$\bar 3$c phases were prepared from the powder by pressing it and subsequent annealing at $1000^\circ$C during 10 hours. We could not obtain ceramic pellets of the \pnma{2} phase.

Magnetic permeability of the \pnma{2}, \pnma{1}, R$\bar 3$c phases was measured in the 77--300~K temperature range in the AC 2500 Hz magnetic field of 1 Oe at slow heating. The measurements were performed on powder manually pressed into a quartz tube of 2 mm in diameter. This technique results in some uncertainty in the amount of material under investigation. Therefore, the absolute value of the permeability was obtained with some uncertainty, yet we determined the main features of its temperature dependence.

IR reflection spectra of ceramic pellets and the IR transmission spectra of powder samples were obtained using a Fourier-transform spectrometer in the spectral range 50--800 cm$^{-1}$ at room temperature. The reflection spectra were measured in the arrangement where the light falls on a pellet surface near perpendicularly, and an aluminum mirror was used to obtain a reference spectrum. In order to measure transmission spectra, either a polyethylene or a KBr plate (depending on the spectral range) was covered by powder sample, and the transmission spectrum of the plate was used as a reference. Transmission $T$ then was converted to absorbance $D=-\ln(T)$.

\section{Results and discussion}

According to X-ray analysis, the unit cell parameters of the \pnma{2}, \pnma{1}, R$\bar 3$c phases we synthesized were the following:\\
\begin{tabular}{c@{\hskip 2em}c@{\hskip 2em}c@{\hskip 2em}c}
phase    & $a$, \AA & $b$, \AA  & $c$, \AA\\
\pnma{1} & 5.505    &  7.776    & 5.513\\
\pnma{2} & 5.732    &  7.693    & 5.536\\
\end{tabular}\\
For R$\bar 3$c $a^*=5.515$~\AA, $c^*=13.291$~\AA\ in the hexagonal coordinates. 

These parameters are concordant, for instance, with the results of Huang \textit{et al.}\cite{Huang}

Titration has shown the following percentage of Mn$^{+4}$ ions in investigated samples: \pnma{2}, 5\%; \pnma{1}, 15\%; R$\bar 3$c 30\%. It corresponds to $\delta$ equal to 0.025, 0.075 and 0.15 for the \pnma{2}, \pnma{1} and R$\bar 3$c phases, respectively.

\begin{figure}
\includegraphics[width=\textwidth,height=\textwidth,keepaspectratio=true]{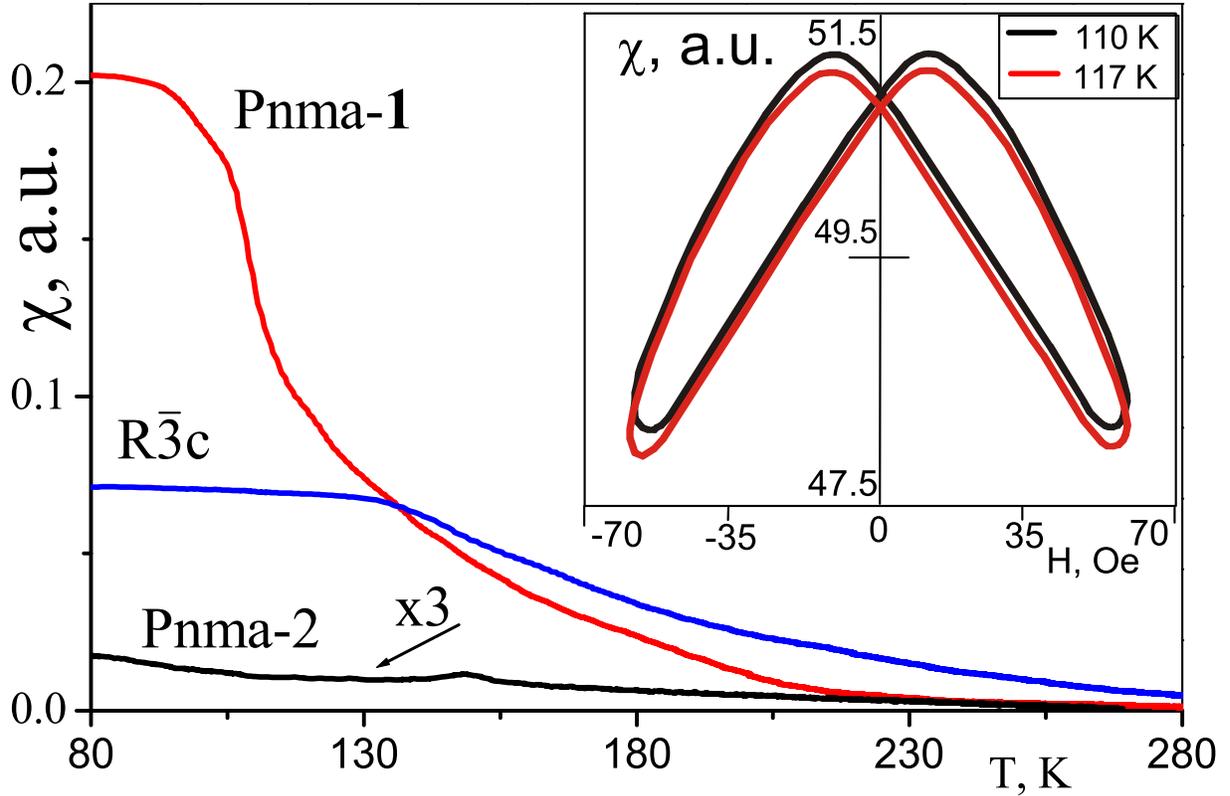}
\caption{\label{fig:fig2} Temperature dependence of the magnetic permeability $\chi(T)$ of the \pnma{2} phase (black, multiplied by 3), the \pnma{1} phase (red) and the R$\bar 3$c phase (blue). For the \pnma{1} phase, permeability versus magnetic field $\chi(H)$ is plotted in the inset at 110 and 117~K.}
\end{figure}

The magnetic permeabilities of \pnma{2}, \pnma{1} and R$\bar 3$c are shown in Fig. \ref{fig:fig2}. All phases are paramagnetic near the room temperature. At low temperature \pnma{1} and R$\bar 3$c are ferromagnetic, and \pnma{2} is antiferromagnetic. Ferromagnetic behaviour is illustrated by hysteretic dependence of the permeability upon the magnetic field, which appears below the transition temperature and becomes more and more pronounced with decreasing temperature, see the inset in Fig. \ref{fig:fig2}. The temperature of the antiferromagnetic transition in \pnma{2} is 140~K, in agreement with Refs. \onlinecite{Huang, Topfer}. To obtain the temperatures of the ferromagnetic transitions in \pnma{1} and R$\bar 3$c, we plotted inverse permeability versus temperature, and linearly extrapolated to zero value the high-temperature parts of these dependences. In agreement with Ref. \onlinecite{Topfer}, the transition temperatures turned out to be 180 and 240~K in the \pnma{1} and R$\bar 3$c phases, respectively. These results confirm that we really deal with the \pnma{2}, \pnma{1} and R$\bar 3$c phases.

\begin{figure}
\includegraphics[width=\textwidth,height=\textwidth,keepaspectratio=true]{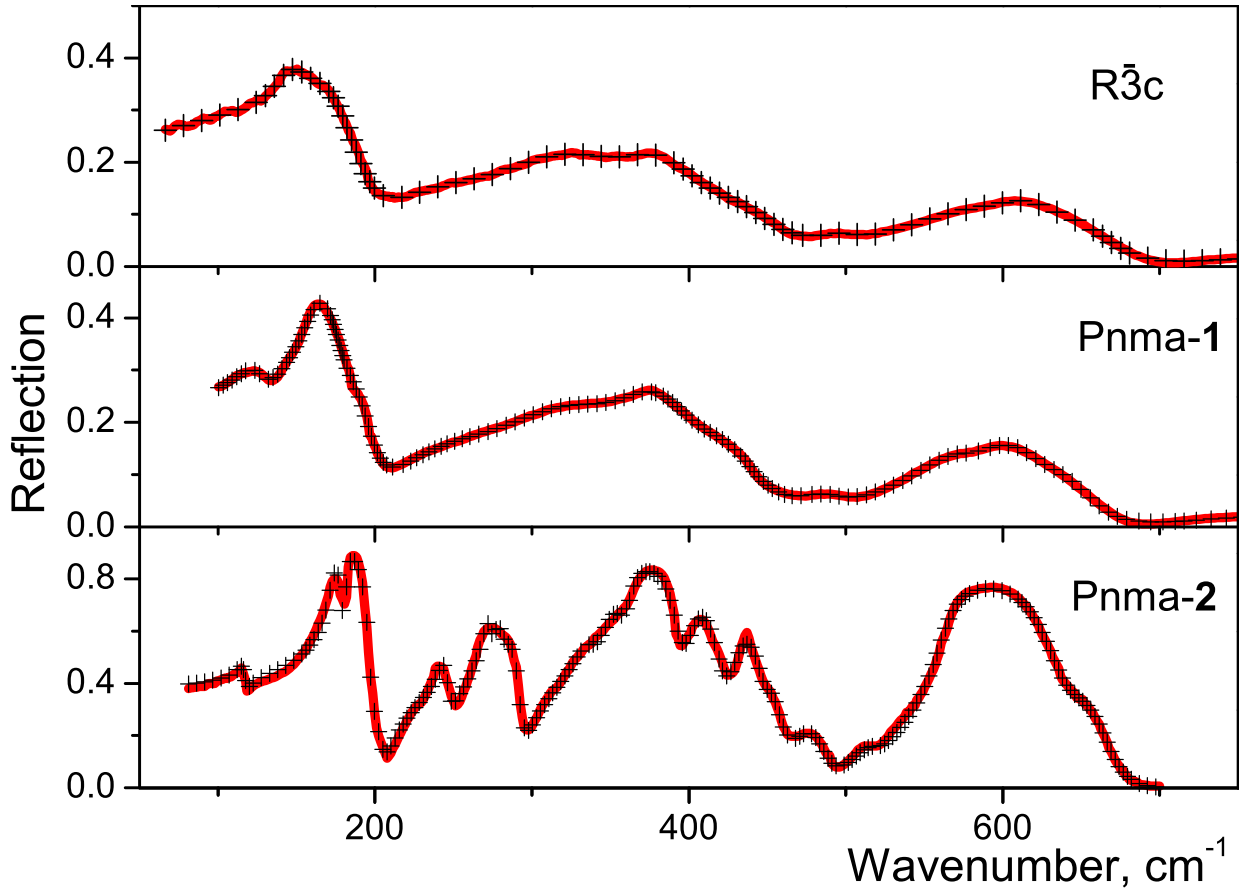}
\caption{\label{fig:fig3} Solid lines: Reflection spectra of the R$\bar 3$c, \pnma{1} and \pnma{2} phases. (For the \pnma{2} phase the data are taken from Ref. \onlinecite{Paolone}). Crosses: the results of fitting.}
\end{figure}

In Fig. \ref{fig:fig3} the reflection spectra of the phases R$\bar 3$c ($\delta \sim 0.15$), \pnma{1} ($\delta \sim 0.05$), and \pnma{2} ($\delta =0$) are shown. In the present wavenumber range reflection spectra are determined by dipole-active phonons. We approximated our reflectivity spectra $R(\omega)$ using a fitting procedure based on a set of Lorentz oscillators:\\
\begin{equation}\label{eq1}
\epsilon(\omega)=\sum_j{\frac{S_j\omega_{0,j}^2}{\omega_{0,j}^2-\omega^2-i\gamma_j\omega}};\quad
R(\omega)=\biggl|{\frac{(\sqrt{\epsilon(\omega)}-1}{\sqrt{\epsilon(\omega)}+1}\biggr|}^2
\end{equation}
$\epsilon(\omega)$ is the complex dielectric function; $S_j, \omega_{0,j}$ and $\gamma_j$ are oscillator strength, frequency and damping factor of mode $j$. The number of oscillators we used in every case was chosen as the minimum number allowing a good fit. The crosses on Fig. \ref{fig:fig3} show the result of the fitting.

\begin{figure}
\includegraphics[width=\textwidth,height=\textwidth,keepaspectratio=true]{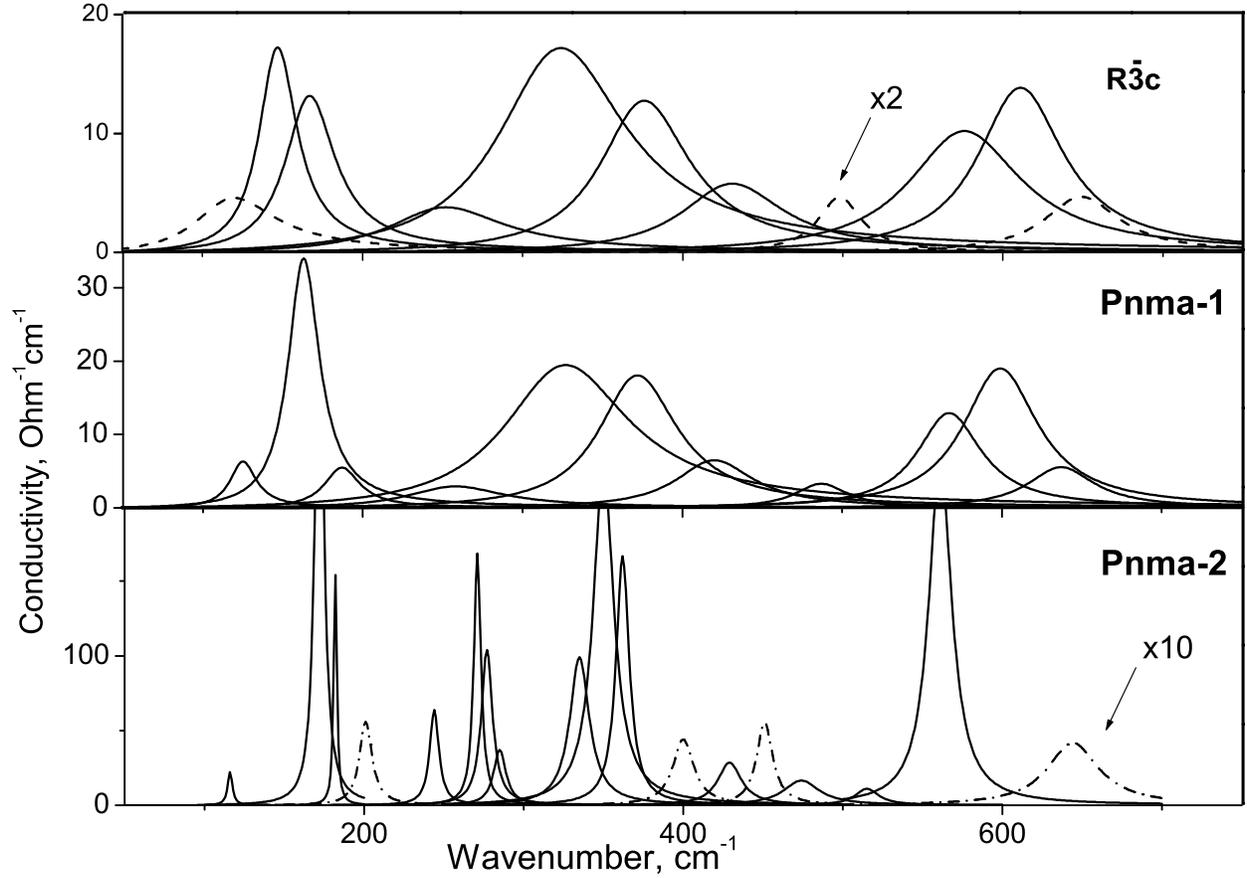}
\caption{\label{fig:fig4} Separate conductivity contributions of each Lorentz oscillator, which were obtained by fitting of the reflection spectra shown in Fig. \ref{fig:fig3}}
\end{figure}

Fig. \ref{fig:fig4} shows the conductivity contributions $\sigma_j(\omega)$ of the calculated Lorentz oscillators:
\begin{equation}\label{eq2}
\sigma_j(\omega)=\frac{1}{4\pi}\cdot\frac{\omega^2\gamma_jS_j}{(\omega_{0,j}^2-\omega^2)^2+\gamma_j^2\omega^2}
\end{equation}

Paolone \textit{et al.}\cite{Paolone} compared experimental and theoretically calculated\cite{Smir} phonon frequencies of \pnma{2} phase. Taking into account the lowest and the highest phonon frequencies obtained by Paolone \textit{et al.}\cite{Paolone}, we corrected previously calculated\cite{Smir} phonon frequencies of the \pnma{2} phase. Also, we calculated the phonon frequencies of the \pnma{1} phase using the rigid-ion model with effective charges. Table \ref{tab:tab1} shows the results of these calculations along with the phonon frequencies extracted from experimental data. We measured spectra of ceramic samples. So the polarization symmetry of the IR-active phonons could not be obtained from our experiments and the arrangement of the modes is tentatively done according to their frequencies and intensities. 

%\squeezetable
\begin{table}
\caption{\label{tab:tab1} Calculated and experimental $TO(LO)$ frequencies (cm$^{-1}$) of IR-active phonon modes; \textit{w} means a weak mode; $\gamma$ is damping factor (cm$^{-1}$)}
\begin{ruledtabular}
\begin{tabular}{llllllll}
\multicolumn{3}{c}{\pnma{2}} &\multicolumn{3}{c}{\pnma{1}}&\multicolumn{2}{c}{R$\bar 3$c}\\
\hline
\multicolumn{1}{c}{calc.}&\multicolumn{2}{c}{exp.}  &\multicolumn{1}{c}{calc.}& \multicolumn{2}{c}{exp.}  & \multicolumn{2}{c}{exp.} \\
\multicolumn{1}{c}{$\omega_{TO}$($\omega_{LO}$)}&\multicolumn{1}{c}{$\omega_{TO}$($\omega_{LO}$)}&\multicolumn{1}{c}{$\gamma$}&\multicolumn{1}{c}{$\omega_{TO}$($\omega_{LO}$)}& \multicolumn{1}{c}{$\omega_{TO}$($\omega_{LO}$)}&\multicolumn{1}{c}{$\gamma$}& \multicolumn{1}{c}{$\omega_{TO}$($\omega_{LO}$)}&\multicolumn{1}{c}{$\gamma$}  \\
%&  & &\multicolumn{1}{c}{$\gamma$}& &  &\multicolumn{1}{c}{$\gamma$}& &\multicolumn{1}{c}{$\gamma$}  \\
\hline
   115(119) $B_{1u}$ &  116(120)& 4    &111(115) $B_{1u}$ &             &     &      &\\
   116(118) $B_{3u}$ &          &       &120(130) $B_{3u}$ &   125(135)  & 20  &120(140)&62\\
   138(140) $B_{2u}$ &          &       &143(148) $B_{2u}$ &             &     &147(180)&29\\
   171(197) $B_{2u}$ &  172(244)&   6   &166(196) $B_{1u}$ &    163(209) & 24  &167(197)&38\\
   175(195) $B_{1u}$ &  182(195)&   3   & 181(199) $B_{2u}$&    187(195) & 27  &&\\
   231(232) $B_{3u}$ &  201(203)&   9   &229(230) $B_{3u}$\textit{w}&             &     && \\
   233(249) $B_{1u}$ &  244(255)&   7   &247(248) $B_{1u}$\textit{w}&             &     &&\\
   249(250) $B_{2u}$ &          &       &300(302) $B_{2u}$\textit{w}&             &     &&\\
   254(281) $B_{3u}$ &  271(291)&   5   & 253(253) $B_{3u}$\textit{w}&            &     && \\
   284(296) $B_{1u}$ &  277(297)&   9   &270(291) $B_{3u}$&     258(267)& 74  &252(266)&88\\
   297(305) $B_{3u}$ &  285(293)&   9   &280(281) $B_{1u}$\textit{w}&             &     &&\\
   309(309) $B_{1u}$ &          &       &332(354) $B_{1u}$&     327(381)& 95  &324(376)&97\\
   330(341) $B_{2u}$ &  335(363)&  15   &355(371) $B_{1u}$&             &     &        &\\
   346(352) $B_{1u}$ & 350(411) &  16   &368(370) $B_{2u}$\textit{w}&             &     &        &\\
   354(373) $B_{3u}$ &  362(391)&  10   &377(440) $B_{3u}$&     372(401)&  60 &376(400)&68 \\
   420(426) $B_{2u}$ &  400(401)&  16   &382(448) $B_{1u}$&             &     &        &\\
   434(450) $B_{1u}$ & 429(437) &  18   &416(417) $B_{1u}$\textit{w}&    420(429) & 59  &431(442)&78\\
   455(457) $B_{1u}$ & 451(452) &  12   &437(444) $B_{3u}$&             &     && \\
   473(479) $B_{3u}$ &  474(480)&  28   &487(503) $B_{2u}$&     487(490)& 40  &498(592)&33\\
   528(531) $B_{3u}$ &  515(518)&  18   &564(568) $B_{2u}$&             &     &&\\
   573(598) $B_{2u}$ &  561(606)&  17   &580(589) $B_{3u}$&      567(579)& 49  &576(592)&85 \\
   634(640) $B_{2u}$ &  644(646)&  39   &584(641) $B_{2u}$&     599(618)& 57  &611(627)&65\\
   644(650) $B_{3u}$ &          &       &615(616) $B_{3u}$\textit{w}&            &     &&\\
   645(651) $B_{1u}$ &         &       &634(639) $B_{1u}$&     637(642)& 51  &649(653)&57\\
\end{tabular}
\end{ruledtabular}
\end{table}

In Table \ref{tab:tab1}, ``$TO$" and ``$LO$" indices correspond to the ``transverse" and ``longitudinal" frequencies. A $TO$ frequency means a resonant frequency $\omega_{0,j}$ (see Equation \eqref{eq1}) and coincides with a maximum of $\sigma(\omega)$ (see Equation \eqref{eq2}). $LO$ frequencies in Table \ref{tab:tab1} correspond to maxima of the function $-\mathrm{Im}(1/\epsilon)$ and represent oscillator strengths $S=\omega_{LO}^2-\omega_{TO}^2$.

\subsection{IR spectra of the Pnma phases}

According to group theory, the isostructural \pnma{1} and \pnma{2} phases should have 25 dipole-active optical phonon modes, 9B$_{1u}$+7B$_{2u}$+9B$_{3u}$ (see, for example, Ref. \onlinecite{Smir}). Indeed, Paolone \textit{et al.}\cite{Paolone} experimentally found 25 IR-active modes in \pnma{2} crystals at 10~K (and 18 modes at room temperature). However, in our \pnma{1} ceramic only 11 modes can be distinguished at room temperature. 

The lines in the \pnma{1} ceramic are substantially wider than in the \pnma{2} single crystals (see damping factors $\gamma$ in Table \ref{tab:tab1}). Let's consider possible reasons for this broadening.

\begin{figure}
\includegraphics[width=\textwidth,height=\textwidth,keepaspectratio=true]{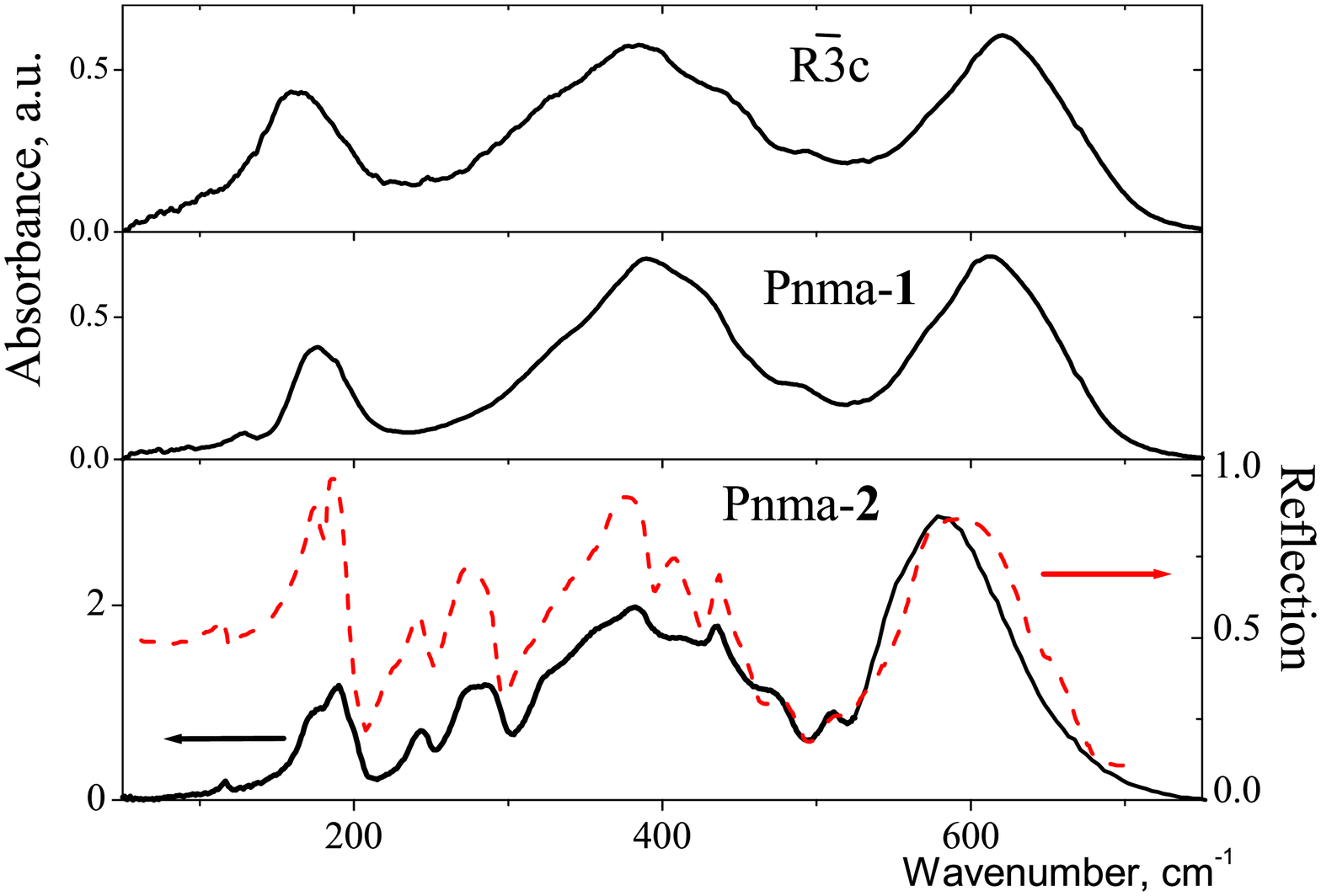}
\caption{\label{fig:fig5} Experimental absorption of the R$\bar 3$c (top), \pnma{1} (middle) and \pnma{2} (bottom, solid line) powders. The dashed line in the bottom part represents the reflectivity of a \pnma{2} single crystal taken from Ref. \onlinecite{Paolone}.}
\end{figure}

Decreasing of the phonon life time $\tau$ accompanied by increasing of $\gamma=1/\tau$ could come as a result of the phonon scattering on grain boundaries of ceramic. To check that, we measured transmission spectra of the \pnma{1}, \pnma{2} and R$\bar 3$c powders. The grain sizes of our powders were measured\cite{Bred} using electron microscopy: in all samples the typical grain size is found to be about 1 $\mu$m. In the transmission spectra, the widths of the phonon lines increase monotonically with the increase of the excess oxygen content, i. e., in the sequence \pnma{2}, \pnma{1}, R$\bar 3$c.  That means that phonon scattering on grain boundaries is not the main reason of line broadening in the spectra of the \pnma{1}, R$\bar 3$c powders. The same is even truer for the spectra of the \pnma{1}, R$\bar 3$c ceramics, because in a ceramic the typical grains can be larger than that in a source powder. Moreover, we believe that even in our \pnma{2} powder phonon scattering on grain boundaries is not the main reason of the line broadening. In the bottom part of Fig. \ref{fig:fig5}, the dashed line shows the reflection spectrum of a \pnma{2} single crystal\cite{Paolone}, solid line represents our absorption spectrum of the \pnma{2} powder. Our calculations showed that, on average, the lines in the conductivity spectrum of powder are three times wider than those in the spectrum of a crystal. Nevertheless, one can reveal the same number of lines in both spectra. For example, 172 cm$^{-1}$ and 182 cm$^{-1}$ lines can be undoubtedly distinguished in our powder spectrum. It was shown\cite{Paolone} that in a doped LaMnO$_{3}$ single crystal, containing 8\% of Mn$^{+4}$, these lines could not be resolved at room temperature. Our powder contained 5\% of Mn$^{+4}$ so it seems reasonable to attribute the observed broadening of lines in our \pnma{2} powder as a result of oxygen doping.

The main factor of line broadening in the spectra of these samples should be the phonon scattering on structural defects, which multiply with excess oxygen doping. These defects could be oxygen atoms in interstitial sites, like those in La$_2$CuO$_{4+\delta}$ [\onlinecite{Chaillout}]. However as for LaMnO$_{3+\delta}$ and La$_{1-x}$A$_{x}$MnO$_{3+\delta}$ (A=Ca, Sr, Ba), at the moment it is rather believed that the nonstoichiometric oxygen O$_\delta$ is compensated by both La and Mn vacancies in equal amounts.\cite{Roosmal,Topfer} In such a case, vacancy contents of La or Mn in our samples $\delta/(3+\delta)$ would be 0.7\%, 2.6\% and 5\% for the \pnma{2}, \pnma{1} and R$\bar 3$c phases respectively.

Line broadening can make difficult or impossible experimental detection of some lines with small oscillator strength. In the \pnma{2} phase, that could be the phonons with the frequencies 400 cm$^{-1}$, 451 cm$^{-1}$ (see Table \ref{tab:tab1}). We calculated the oscillator strength for all IR-active modes of the \pnma{2} and \pnma{1} phases. It turns out that the number of modes experimentally detected in the \pnma{1} phase is reduced in comparison with the \pnma{2} phase mainly because the oscillator strength of some phonons of the \pnma{1} phase becomes very small. These \pnma{1} modes are marked by \textit{w} in Table \ref{tab:tab1}.  In the \pnma{1} phase, the lengths of Mn--O bonds differ from each other very little (the difference comes in fourth significant digit). The closeness of Mn--O bond lengths means that oxygen atoms are almost symmetrically equivalent, i. e., the \pnma{1} crystal structure deviates from the cubic one less than the \pnma{2} crystal structure where the difference in Mn--O bond lengths is 15\%. In the cubic structure, the number of IR-active phonons is less than in an orthorhombic structure. Therefore, if a structure is close to cubic then some IR-active phonons are ``on the verge of disappearance".

\subsection{IR spectra of R$\bar 3$c}

Our spectra of R$\bar 3$c are in satisfactory agreement with the spectra obtained in Ref. \onlinecite{Abrashev,Kebin}.

According to our experimental results, phonon damping factors of the R$\bar 3$c phase exceed those of the \pnma{1} phase by a factor of 1.3 on average. The first reason is that the Mn$^{+4}$ content in R$\bar 3$c is two times as large as it is in the \pnma{1} phase, so there are more structural defects there. The second reason is disorder caused by the noncoherent dynamic Jahn-Teller effect.

According to the group-theory analysis (see Ref. \onlinecite{Smir}, for example), there are 8 IR-active phonon modes in the R$\bar 3$c phase: 3A$_{2u}$+5E$_u$. At room temperature, in reflection spectra of the R$\bar 3$c ceramic we definitely distinguish 10 lines. The approximation by a set of Lorentz oscillators revealed an additional very broad line near 120 cm$^{-1}$. Therefore, we found in the R$\bar 3$c phase the same amount of lines (11) as in the \pnma{1} phase. 

Let us consider possible reasons for appearing of additional lines in spectra of the R$\bar 3$c phase.

Local break of the inversion symmetry around a point defect could make some Raman-active (IR-forbidden) modes to appear in IR spectra. However, comparison of the IR spectra of the R$\bar 3$c phase with Raman spectra of Abrashev \textit{et al.}\cite{Abrashev} shows that there is only one Raman line near 649 cm$^{-1}$ close to an IR line (640 cm$^{-1}$), the other Raman lines have no counterparts in our IR spectra.

In IR spectra there could appear maxima of the phonon density of states caused by breaking of the long-range order. Iliev \textit{et al.}\cite{Iliev2} analyzed the Raman spectra of doped rare-earth manganites and interpreted them in the frame of the model used for description of amorphous materials.\cite{Shuker} The Raman spectra in this case are dominated by disorder-induced bands, reflecting the phonon density of states smeared due to finite phonon lifetime. In other words, the law of conservation of the quasimomentum $\mathbf{k}$ breaks and phonons with nonzero $k$ begin to interact with light. In general, the same mechanism could definitely work for IR spectra too. Big linewidths prevent us from supporting or rejecting an influence of phonons with $k\neq 0$ on IR spectra of the R$\bar 3$c phase. Though it worth to take into account that according to Iliev \textit{et al.}\cite{Iliev2} a Raman mode generally gives several maxima of density of states. Probably the same is true for IR-active modes. However, our spectra of the R$\bar 3$c phase can be fitted very well by a few Lorentz functions. So we think that the phonons with $k\neq 0$ can have only a small influence on our spectra, they do not determine essential spectral features.

We explain additional lines in our IR spectra of the R$\bar 3$c phase as a result of the dynamic Jahn-Teller effect. In the R$\bar 3$c phase of LaMnO$_{3}$, the R$\bar 3$c symmetry exists only ``on average", revealing itself in certain kinds of experiments such as X-ray diffraction. At any particular moment of time, one of the octahedron O--Mn--O axes differs from two others due to dynamic Jahn-Teller distortions; therefore, oxygen atoms are inequivalent and their charges are not equal. It is the ``instant", not ``average", pattern that is probed in optical experiments.\cite{Iliev2} Obviously, normal phonon modes, measured by means of IR and Raman spectroscopy, are normal modes of the ``instant", not average" pattern. In the ``instant" view every octahedron in the R$\bar 3$c phase looks deformed, mostly in the same way as the octahedra in the Pnma phases. That's why the phonon spectrum of the R$\bar 3$c phase resembles that of the Pnma phases. Similarly, Abrashev \textit{et al.}\cite{Abrashev} interpreted two strongest lines (649 cm$^{-1}$ is one of them) in their Raman spectra of the R$\bar 3$c phase as ``forbidden" modes, analogous to the respective modes in Pnma phases.

We can expect some correlations between the Jahn-Teller deformations of the octahedra in the R$\bar 3$c phase. Qiu \textit{et al.}\cite{Qiu} found that in high-temperature ($T>1010$~K) stoichiometric rhombohedral LaMnO$_3$ there are fully distorted MnO$_6$ octahedra, ordered in clusters of diameter $\sim 16$ \AA. According Ref. \onlinecite{Topfer}, the phase diagram of LaMnO$_{3+\delta}$ containes an area ($0.11<\delta<0.14$) where a phase transition R$\bar 3$c $\Leftrightarrow$ \pnma{1} exists at $T = 300$~K. As we mentioned in Section II, such transition of a second kind is forbidden by symmetry. In Ref. \onlinecite{Shehtman} there was suggested a model of a phase transition through a virtual cubic phase. Taking into account the known IR and Raman spectra of the R$\bar 3$c phase, as well as the results of Qiu \textit{et al.}\cite{Qiu}, we suggest that the R$\bar 3$c samples could contain nanoclusters of some Pnma phase. Such inclusions may be growing centres at the transition R$\bar 3$c $\Leftrightarrow$ \pnma{1} of a first kind.

\section{The influence of selection rules of D$_{2h}$ point group on the IR spectra}

According to the selection rules, the irreducible representations B$_{1u}$, B$_{2u}$, B$_{3u}$ of D$_{2h}$ point group correspond to IR-active modes, their total electric dipole moment $\bm{M}$ taking the form $\bm{M}(\textrm{B}_{1u})=(0,0,M_z), \bm{M}(\textrm{B}_{2u})=(0,M_y,0), \bm{M}(\textrm{B}_{3u})=(M_x,0,0)$. Similarly, for every full set of symmetrically equivalent atoms in the unit cell (O2, for example) the sum of their atomic displacements $\sum_i{\bm{u}_i}$ has only one non-zero component. (For a single atom inside such a set, all three components can differ from zero.)

\begin{figure}
\includegraphics[width=\textwidth,height=\textwidth,keepaspectratio=true]{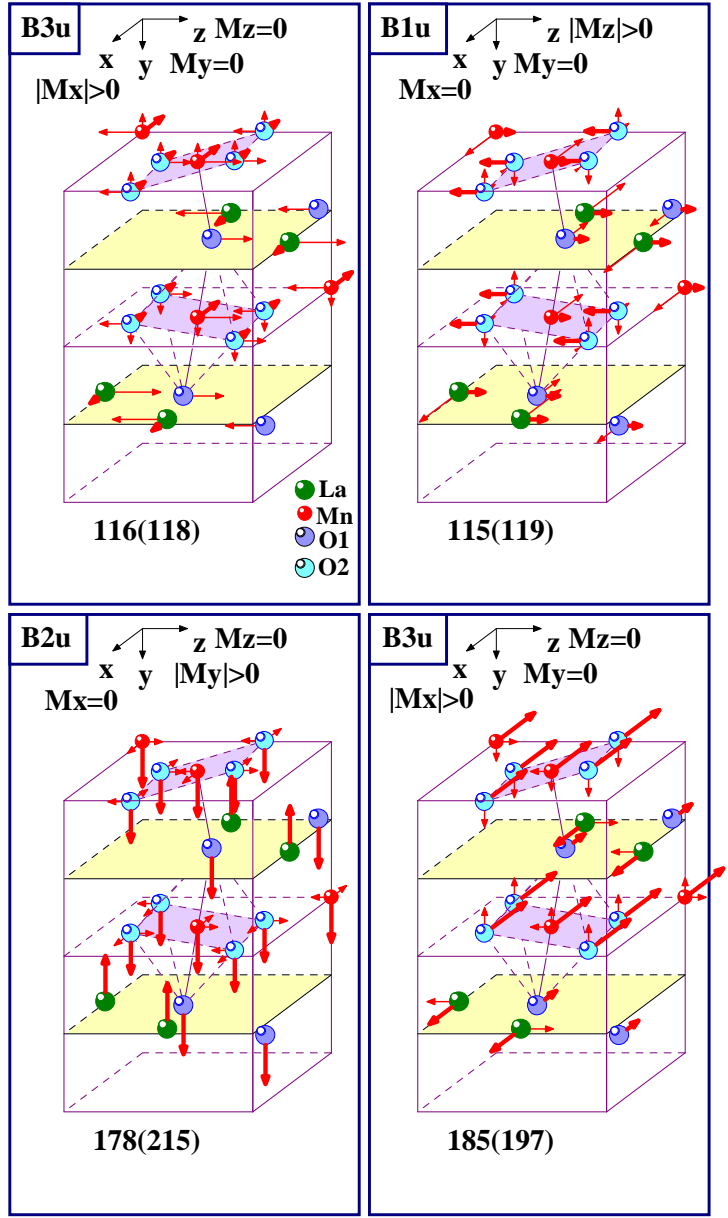}
\caption{\label{fig:fig6} Theoretically calculated patterns of some IR-active phonon modes of the \pnma{2} phase. Thick arrows show atomic displacements in the direction of the total electric dipole moment $\bm{M}$. Thin arrows show atomic displacements in the other two main crystallographic directions. In the left upper corners there shown corresponding irreducible representations. In the bottom there shown corresponding theoretical TO(LO) frequencies.}
\end{figure}

Let's consider four lowest-frequency IR-active modes of the \pnma{2} phase. (Fig. \ref{fig:fig6})

The line with the lowest frequency (115 cm$^{-1}$) can be distinctly seen in the spectra of the \pnma{2} and \pnma{1} phases. In the spectrum of the R$\bar 3$c phase it substantially broadens (Fig. \ref{fig:fig4}). A similar line have been observed in reflection spectra of both the undoped ( $x=0$ ) and doped by either Ca or Sr La$_{1-x}$A$_x$MnO$_{3+\delta}$, LaTiO$_3$\cite{Lunkenheimer}, YVO$_3$\cite{Tsvetkov}. Theoretical calculations\cite{Smir} and experimental results\cite{Tsvetkov} show that in the spectra of the \pnma{2} phase this line consists of two modes with close frequencies and different polarizations (see the upper part of Fig. \ref{fig:fig6}). 

In B$_{1u}$, B$_{3u}$ modes, La and O1 atoms can vibrate only in the reflection plane $\bm{m}$ therefore having two degrees of freedom. 

115 cm$^{-1}$ mode (B$_{1u}$) has the maximal displacements of La atoms along $x$ axis. Nevertheless, these components don't contribute to the total electrical dipole moment because their sum equals zero. Only small components of the La displacements $\bm{u}_z$ along $z$ axis (thick arrows) contribute to $\bm{M}$. The intensity of this mode in the optical conductivity spectrum is determined by the displacements of O2, Mn, La atoms, their contributions adding together. Relatively small contributions of O1 atomic displacements have the opposite sign.

The structure of atomic displacements of 116 cm$^{-1}$ mode (B$_{3u}$) is similar to the previous one. The biggest displacements of La atoms are along $z$ axis, $\bm{M}$ being parallel to $x$ axis. The intensity of this mode is determined by the adding contributions of O2, La displacements and the subtracting contribution of Mn displacement.

In 178 cm$^{-1}$ mode (B$_{2u}$) O1 and La atoms can vibrate only along $y$ axis, in 185 cm$^{-1}$ mode (B$_{3u}$) they can vibrate only in (0,1,0) plane. An essential difference between these modes and 115 cm$^{-1}$, 116 cm$^{-1}$ modes is that in 178 cm$^{-1}$, 185 cm$^{-1}$ modes the maximal displacements of every atom contribute to $\bm{M}$ (O1, O2, La are adding, Mn is subtracting). That is why the oscillator strengths of 178 cm$^{-1}$, 185 cm$^{-1}$ modes are much higher than that of 115 cm$^{-1}$, 116 cm$^{-1}$ modes.

\begin{figure}
\includegraphics[width=\textwidth,height=\textwidth,keepaspectratio=true]{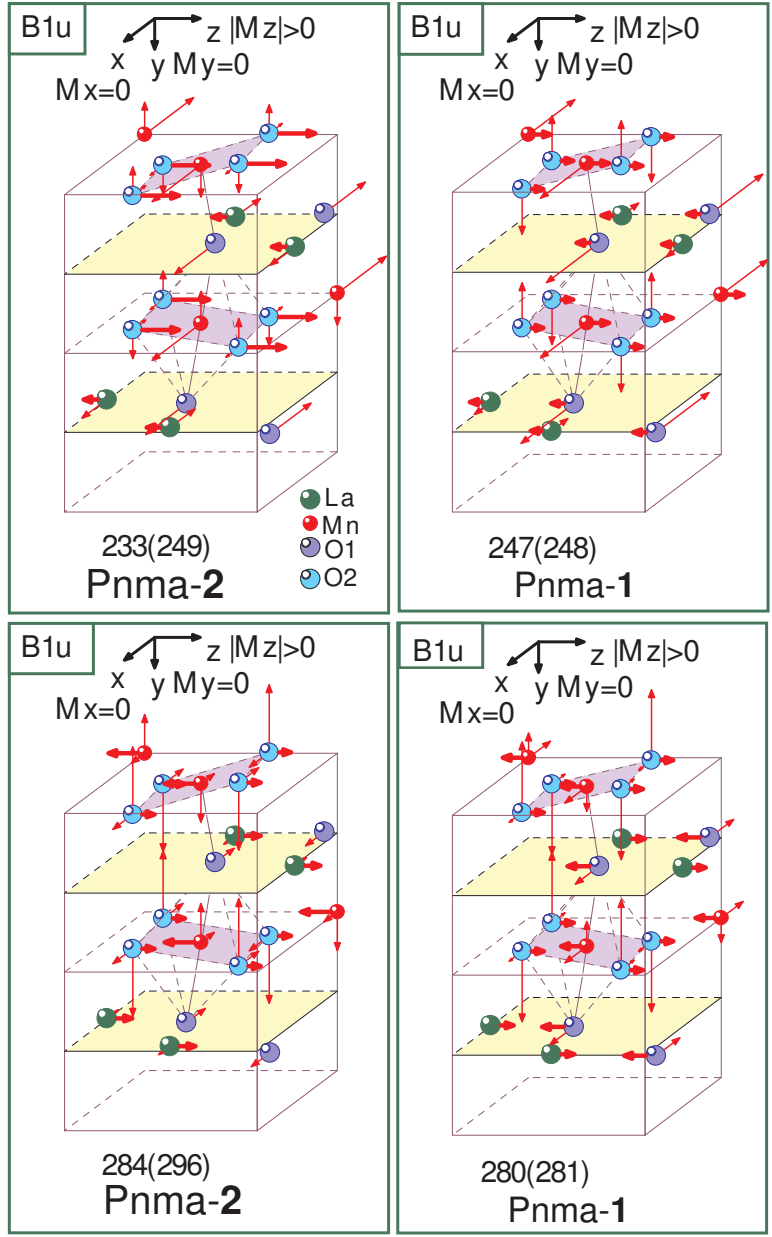}
\caption{\label{fig:fig7} Theoretically calculated patterns of some IR-active phonon modes of the \pnma{2} and \pnma{1} phases. Thick arrows show atomic displacements in the direction of the total electric dipole moment $\bm{M}$. Thin arrows show atomic displacements in the other two main crystallographic directions. In the left upper corners there are shown corresponding irreducible representations. In the bottom there shown corresponding theoretical TO(LO) frequencies.}
\end{figure}

Being isostructural, the \pnma{2} and \pnma{1} phases have close patterns of atomic displacements in phonon modes. Still, there are some important differences between them. In the upper part of Fig. \ref{fig:fig7} there are shown 233 cm$^{-1}$ mode of the \pnma{2} phase and 247 cm$^{-1}$ mode of the \pnma{1} phase. Big displacements of Mn and O1 along $x$ axis, which have comparable magnitudes for the \pnma{2} and \pnma{1} phases, don't contribute to $\bm{M}$. In the both cases, the oscillator strengths are entirely determined by small displacements along $z$ axis, which are much less for the \pnma{1} phase (247 cm$^{-1}$) than for the \pnma{2} phase (233 cm$^{-1}$). As a result, the oscillator strength 247 cm$^{-1}$ mode of the \pnma{1} phase is very small.

In the bottom part of Fig. \ref{fig:fig7} there are shown another pair of similar modes. The oscillator strength of 284 cm$^{-1}$ mode (\pnma{2}) is much higher than that of 280 cm$^{-1}$ mode (\pnma{1}), because in the second case the displacements of Mn, O2 atoms along $z$ axis are substantially less. In addition, the displacements of O1 atoms, which decrease the resulting $\bm{M}$, are of much higher amplitude in 280 cm$^{-1}$ mode (\pnma{1}) than in 284 cm$^{-1}$ mode (\pnma{2}).

Our theoretical calculations showed that there are six modes in total, which strongly decrease their oscillator strength for the \pnma{1} phase in comparison with that for the \pnma{2} phase. (In Table \ref{tab:tab1} they are marked by \textit{w}.) That's why for the \pnma{1} phase the number of modes seen in experiment is less than for the \pnma{2} phase.

\begin{figure}
\includegraphics[width=\textwidth,height=\textwidth,keepaspectratio=true]{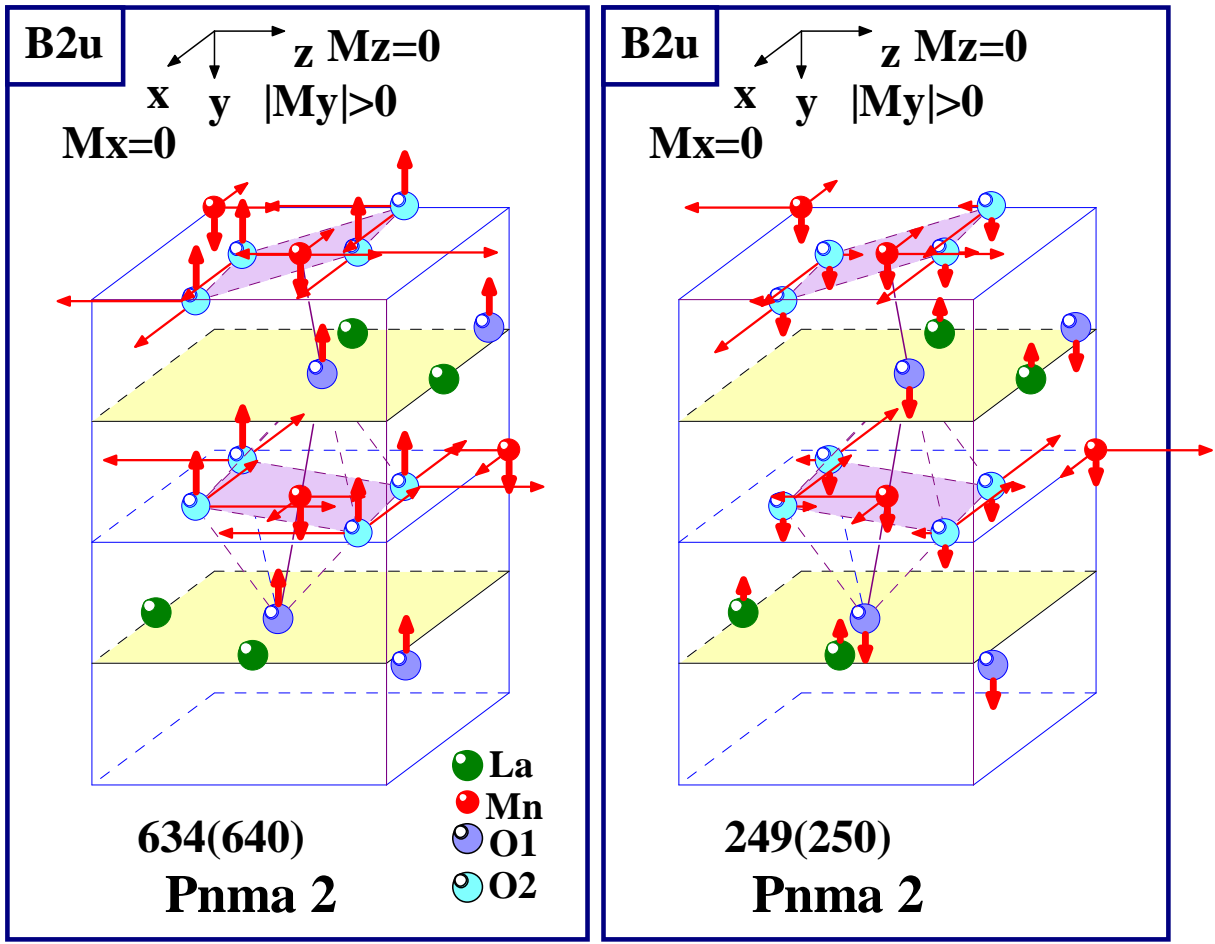}
\caption{\label{fig:fig8} Theoretically calculated patterns of some IR-active phonon modes for the \pnma{2} phase. Thick arrows show atomic displacements in the direction of the total electric dipole moment $\bm{M}$. Thin arrows show atomic displacements in the other two main crystallographic directions. In the left upper corners there are shown corresponding irreducible representations. In the bottom there shown corresponding theoretical TO(LO) frequencies.}
\end{figure}

The atomic displacements of all IR-active modes for the \pnma{2} phase are drawn in Fig. 5 of Ref. \onlinecite{Smir}. Mostly, the displacements of O1, O2 atoms are much bigger than that of Mn, La atoms. As a result, the small components were ignored there. For a strong mode, that was reasonable. However for a weak mode, that could cause some misunderstanding. For example, all the displacements shown in Ref. \onlinecite{Smir} for 207 cm$^{-1}$ and 562 cm$^{-1}$ modes produce the resulting $\bm{M}=0$. More correct patterns for these modes are shown in Fig. \ref{fig:fig8}.

\section{Conclusions}

The reversible sequence of transformations R$\bar 3$c $\Leftrightarrow$ \pnma{1} $\Leftrightarrow$ \pnma{2} was realized by annealing of LaMnO$_{3+\delta}$ powder at $600^\circ$C during 5--10 hours.

For the first time, IR transmission and reflection spectra of the \pnma{1} phase of LaMnO$_{3+\delta}$ were measured. In addition, IR spectra of the \pnma{2} and R$\bar 3$c phases were measured and found to be in satisfactory agreement with previously published results.

Taking into account new experimental data for the \pnma{2} phase, we corrected our parameters of the rigid-ion model with effective charges and recalculated its phonon spectrum. The frequencies and oscillator strengths of the IR-active phonons in \pnma{1} phase were calculated as well.

The number of experimentally observed IR-active phonon modes in the \pnma{1} phase is smaller than that in the \pnma{2} phase, although these phases have the same Pnma symmetry.
According to theoretical calculations, it happens due to a decrease in the oscillator strengths of several phonon modes of the \pnma{1} phase. The underlying reason is that in the \pnma{1} phase MnO$_6$ octahedra are much less distorted than in the \pnma{2} phase.

In the spectra of the R$\bar 3$c phase, the number of modes observed exceeds that predicted by group theory. We attribute the additional modes to local distortions of oxygen octahedra similar to those in Pnma phases.

\begin{acknowledgments}
We thank S.~S.~Nazin for useful discussion.
\end{acknowledgments} 

\bibliography{lamnO3h}

\end{document}